\documentclass{article}
\usepackage{sonihed}
\usepackage{times}
\usepackage{ifpdf}
\usepackage[english]{babel}
\usepackage{cite}
\usepackage{url}

\def\papertitle{Interactive Sonification for Health and Energy using ChucK and Unity}
\def\firstauthor{Yichun Zhao}
\def\secondauthor{George Tzanetakis}

\newif\ifpdf
\ifx\pdfoutput\relax
\else
   \ifcase\pdfoutput
      \pdffalse
   \else
      \pdftrue
\fi

\ifpdf 
  \usepackage[pdftex,
    pdftitle={\papertitle},
    pdfauthor={\firstauthor, \secondauthor}, 
    bookmarksnumbered, 
    pdfstartview=XYZ 
   ]{hyperref}
  \usepackage[pdftex]{graphicx}
  \graphicspath{{./figures/}}
  \DeclareGraphicsExtensions{.pdf,.jpeg,.png}
  \usepackage[figure,table]{hypcap}

\else 
  \usepackage[dvips,
    bookmarksnumbered, 
    pdfstartview=XYZ 
  ]{hyperref}  
  \usepackage[dvips]{epsfig,graphicx}
  \graphicspath{{./figures/}}
  \DeclareGraphicsExtensions{.eps}
  \usepackage[figure,table]{hypcap}
\fi

\hypersetup{
    colorlinks,%
    citecolor=black,%
    filecolor=black,%
    linkcolor=black,%
    urlcolor=black
}

\title{\papertitle}

\twoauthors
  {\firstauthor} {University of Victoria \\ %
    {\tt \href{mailto:yichunzhao@uvic.ca}{yichunzhao@uvic.ca}}}
  {\secondauthor} {University of Victoria \\ %
    {\tt \href{mailto:gtzan@ieee.org}{gtzan@ieee.org}}}

\begin{document}
\capstartfalse
\maketitle
\capstarttrue
\begin{abstract} 
\sloppy
Sonification can provide valuable insights about data but most existing approaches are not designed to be controlled by the user in an interactive fashion. Interactions enable the designer of the sonification to more rapidly experiment with sound design and allow the sonification to be modified in real-time by interacting with various control parameters. In this paper, we describe two case studies of interactive sonification that utilize publicly available datasets that have been described recently in the International Conference on Auditory Display (ICAD). They are from the health and energy domains: electroencephalogram (EEG) alpha wave data and air pollutant data consisting of nitrogen dioxide, sulfur dioxide, carbon monoxide, and ozone. We show how these sonfications can be re-created to support interaction utilizing a general interactive sonification framework built using ChucK, Unity, and Chunity. In addition to supporting typical sonification methods that are common in existing sonification toolkits, our framework introduces novel methods such as supporting discrete events, interleaved playback of multiple data streams for comparison, and using frequency modulation (FM) synthesis in terms of one data attribute modulating another. We also describe how these new functionalities can be used to improve the sonification experience of the two datasets we have investigated.
\end{abstract}

\section{Introduction}
\label{sec:intro}



\sloppy
The importance of interactivity in sonification is generally accepted, but many existing data sonifications are still limited in terms of their interactivity and when they are interactive they tend to be singular prototype systems created specifically for a particular dataset. In fact, most of the recent sonifications of health or energy data published in the International Conference on Auditory Display (ICAD) are limited in this sense, as summarized in Section \ref{subsection:prev_sonification}. 

In this paper, we introduce an interactive sonification framework which builds upon existing toolkits and utilizes new functionalities to improve the sonification experience so that it is easier for domain experts with no programming experience to adopt and is reusable, avoiding reinventing the wheel. Moreover, we focus on sonifications of data in the health and energy domains in recent years as case studies to demonstrate the framework.

\section{Related Works}
\label{sec:related}

In this section, we summarize works related to sonification toolkits with graphical user interfaces (GUI) and sonifications of energy and health data from recent years, and organize them in chronological order. We then discuss specifically our contributions in relation to previous works.  

\subsection{Sonification Toolkits with GUI}
\sloppy
The Sonification Sandbox was developed by Walker and Cothran \cite{Walker_Cothran_2003}, allowing mappings between data attributes and timbre, pitch, volume and pan. Percussive context could also be added to the sonification. This was later upgraded by Davison and Walker \cite{Benjamin_K._Davison_Bruce_N._Walker_2007} to rebuild the system to allow integration into other systems and saving descriptions of the sonification representation in various formats. Pauletto and Hunt \cite{Pauletto_Hunt} described an interactive sonification toolkit where sonification methods could be changed rapidly, and the current position in the dataset sonified is controllable. The sonification research group at Berne University of the Arts (BUA) \cite{Sonifier_BUA} developed project Sonifier which supports frequency modulation (FM) and notably sonified electroencephalogram (EEG), magnetic resonance imaging (MRI), and seismological data. 

More recently, the sonification workstation was developed by Phillips and Cabrera \cite{Phillips_Cabrera_2019} to increase the accessibility and ease of use of a general sonification toolkit especially for users who might not be familiar to sonification. Similarly, the WebAudioXML Sonification Toolkit (WAST) built by Lindertorp and Falkenberg \cite{Lindetorp_Falkenberg_2021} aims for the same goal but in the web browser. WAST was also evaluated with user studies. The paper suggested that involving instruments and features from Digital Audio Workstations (DAW) would create a more creative environment which is something that has also informed our research. The company Sonify has also developed a web tool, TwoTone \cite{TwoTone_2022} that allows users to map data to musical pitches with different instruments. 

\subsection{Sonifications of Health and Energy Data}
\label{subsection:prev_sonification}

Works sonifying data related to health from ICAD include CardioSounds which is a portable system developed by Blanco et al. \cite{Blanco_Grautoff_Hermann_2018} to sonify electrocardiogram data and enhance rhythmic details in real-time to diagnose and monitor cardiac pathologies. Winters et al. \cite{Winters_Kalra_Walker_2019} sonified the internal workings of artificial neural networks for melanoma diagnosis such as the activation of nodes, seriousness of the conditions of lesions classified, and cluster centers after dimensionality reduction. A sound art piece \textit{The Alchemy of Chaos} was created by Mitchell et al. \cite{Mitchell_Thom_Pountney_Hyde_2019} to discretely sonify episodes from Touette’s syndrome and captured both the contextual information of the episodes and the aesthetics of sound design. Kantan et al. \cite{Kantan_Spaich_Dahl_2021} developed a domain-specific prototype system that involves interactive ``real-time mapping control and physical modelling-based musical sonification'' to support biofeedback for movement rehabilitation. Frid et al. created sonification of the interactions between a COVID-19 patient and a medical team by using electrocardiogram (ECG) data and mapping between properties of heart signals and harmonic tension and dissonance. 

In terms of data related to energy, Cowden and Dosiek \cite{Cowden_Dosiek_2018} presented auditory representations of power grid voltage data with easy-to-use audification and sonification algorithms. Groß-Vogt et al. \cite{Groß-Vogt_Weger_Höldrich_Hermann_Bovermann_Reichmann_2018} sonified the electric power consumption of an institute's kitchen by altering the reverberation of the kitchen based on consumption data. 


As for the two sonifications that are used as case studies later in this paper, one of them is by Steffert et al. \cite{Steffert_Holland_Mulholland_Dalton_Väljamäe_2015} who used amplitude and frequency modulation to sonify EEG data and proposed an evaluation method. The other is by St Pierre and Droumeva \cite{St_Pierre_Droumeva_2016} who used FM and granular synthesis to sonify air pollutant data with the notion of ``harmonic identities''. They are further described in Section \ref{sec:cases}.


\subsection{Our Work}


\sloppy
Our interactive sonification framework generally builds upon the work on accessible sonification toolkits \cite{Phillips_Cabrera_2019} \cite{Lindetorp_Falkenberg_2021} by utilizing modern technology for both the audio generation and the interactive interface to support portability across platforms and software maintenance. It is created using ChucK \cite{Wang_Cook_2003} (a programming language creating concurrent strongly-timed procedural audio), Unity \cite{Unity_Technologies} (a game engine creating the user interface, and spatializing and visualizing the sonification in this project), and Chunity \cite{Atherton_Wang} (which combines both ChucK and Unity allowing integrated audiovisual programming). In comparison to a domain-specific system \cite{Kantan_Spaich_Dahl_2021}, our work can be used for any dataset in a tabular form. In addition to supporting most functionalities of existing sonification toolkits, we also further support de Campo’s sonification design space \cite{de_Campo_2007} to distinguish between continuous and discrete sonification. Furthermore, we introduce novel methods which existing toolkits do not support, such as interleavingly playing sonifications of multiple data streams for comparison, using FM synthesis in terms of one data attribute modulating another, and supporting synchronized interactive 2D and 3D visualizations.

\section{Interactive Sonification} 
\label{sec:interactive}

\begin{figure}[t]
  \centering
  \includegraphics[width=0.99\columnwidth]{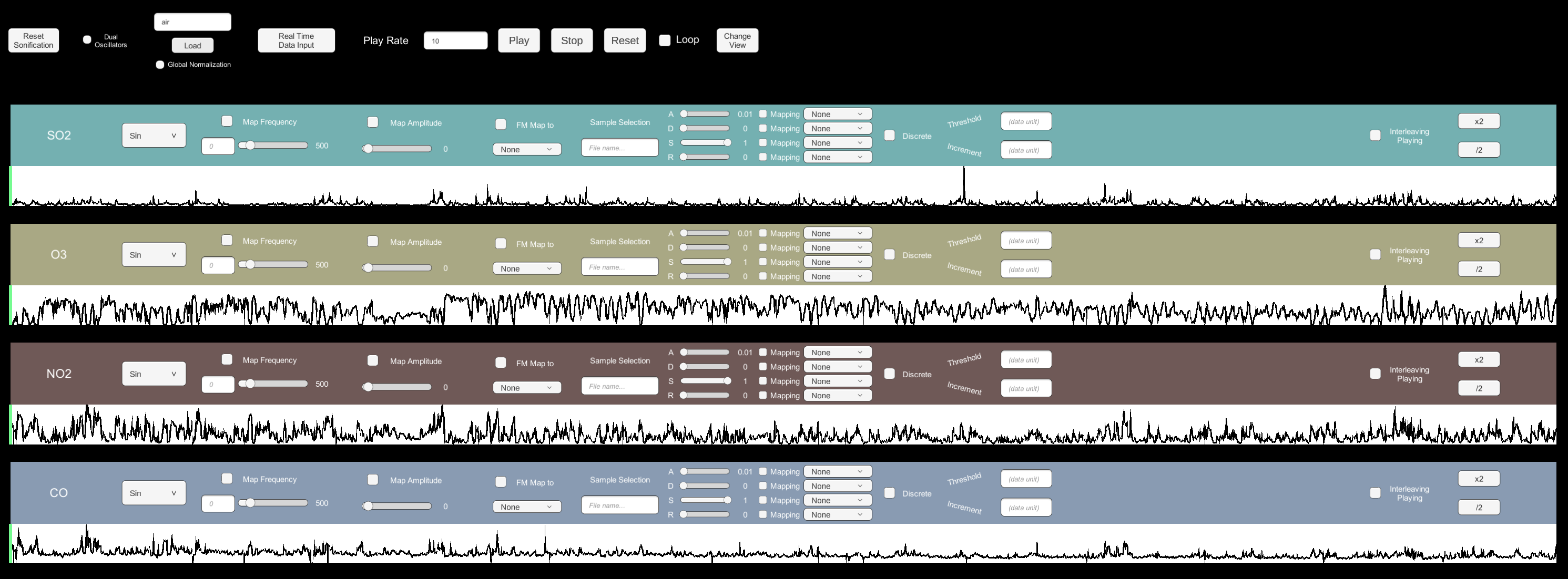}
\caption{The main track layout of the sonification framework when a dataset consisting of 4 data attributes is first loaded with the default configuration.
\label{fig:layout}}
\end{figure}

The general design of our work is inspired by a typical DAW where the top is general controls of data loading (selecting datasets and normalization type), audio playing (play rate, starting, stopping, and resetting), and mode selection (switching to audio visualization). After a dataset is loaded, each data attribute or stream populates a track view, and similar to how a typical DAW looks like, the tracks are horizontal rows stacked together. The top half of a track consists of sound synthesis controls and parameter mappings, and the bottom half is a visual plot of the data stream as a line graph. Each track is randomly and differently coloured not only for aesthetic reasons but also for easy identification of data attributes in the audio visualization view. When the sonification is played, there is a cursor for each track on the visual plot to signify the location of the current data point being sonified. Lastly, when multiple tracks are populated exceeding the current field of view, the tracks could be scrolled up and down to view on the desired track(s). Figure \ref{fig:layout} shows a sample layout.

\section{Case Studies}
\label{sec:cases}

In this section, we describe the two datasets that we sonify as case studies: EEG and air quality data. They are chosen because they are the most recent publicly available data from ICAD that are related to the health and energy domains. In each subsection of them, we describe the replication of sonification from the original paper, and discuss how our sonification framework achieves more with interactivity and new functionalities that are appropriate to the context of the dataset. 

\subsection{EEG Data Sonification}

The EEG data was recorded under the two conditions of having the subject's eyes being open and eyes being close. The alpha waves were separated into low alpha and high alpha frequency, resulting in 4 sets of data for eyes-closed and eyes-open in the high and low alpha frequency conditions \cite{Steffert_Holland_Mulholland_Dalton_Väljamäe_2015}.

\subsubsection{Replication of Sonification}

\begin{figure}[t]
\centering
\includegraphics[width=0.95\columnwidth]{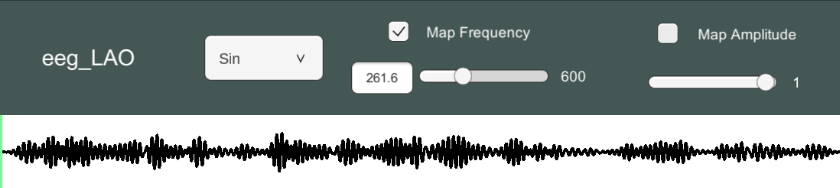}
\caption{Frequency modulation by EEG data with a minimum frequency of 261.6 Hz and a frequency range of 600 Hz.\label{fig:eeg_replicate}}
\end{figure}

After the data is loaded, a minimum frequency can be set in the interactive interface and it is 261.6 or 523.2 Hz from the original paper. The play rate is set to 2 milliseconds per data point to replicate the duration of 1 minute for the sonification. For frequency modulation, the data is configured to be mapped to the frequency of the sine oscillator in the interface; The range of frequencies is set to 600 Hz using a slider as seen in Figure \ref{fig:eeg_replicate}. For amplitude modulation, the data is mapped to the amplitude (the data is automatically scaled to be between 0 and 1 inclusively); The frequency stays at 261.6 or 523.2 Hz. 

\subsubsection{Achieving More}
\label{achieve_more_eeg}


The original paper used Pure Data \cite{Puckette_1996} to create the sonification and needs to load each set of data individually. In our work, all 4 sets could be loaded together and muted or unmuted separately, eliminating the need to reload data when the user wishes to focus on another dataset. Moreover, the sound design parameters and the play rate can be modified in real-time. Although this was partially achieved in the original Pure Data patch, certain controls are easier in our work such as directly setting the frequency range instead of constructing calculations. Changing the oscillator type in Pure Data would require the removal of the oscillator object and re-patching to a different one, but here we could just select another sound option in the drop-down menu for the sound source. 

\begin{figure}[t]
  \centering
  \includegraphics[width=0.65\columnwidth]{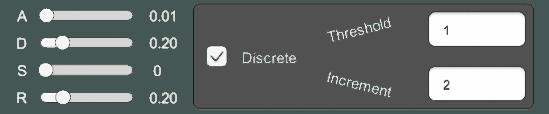}
\caption{The configuration of discrete sonification of the EEG data with envelope parameters modifying the shape of the sound and threshold parameters to specify the conditions for discrete sounds to be played.
\label{fig:eeg_more}}
\end{figure}

The original sonification is only continuous, but our work also supports discrete events. Two parameters ``threshold'' and ``increment'' could be set. Threshold refers to a certain value for the data to reach in order to produce a sound, as demonstrated in \href{https://youtu.be/EHmMB9ddKAU}{this video}\footnote{\url{https://youtu.be/EHmMB9ddKAU}}; Increment refers to a value that could be added to or subtracted from the threshold every time the threshold is reached. An example is shown in Figure \ref{fig:eeg_more} where the threshold is set to 1 and the increment is 2 in this case; A discrete sound is played when the EEG data reaches 1, 3, 5, ..., and so on. (This is especially useful when the data has an increasing or decreasing trend.) Conditional discrete events allow more ways of analyzing EEG data through sonification to only hear a sound when certain conditions are met, instead of hearing the whole audio.  Furthermore, discrete events not only support the use of oscillators but also user-defined audio samples by selecting the "Sample" option in the sound source drop-down menu and specifying the file name of the sample. The audio samples can be further manipulated by changing their speed and amplitude and thus affecting their pitch and volume. This enables any kind of audio to be played as the user desires. 


Envelope controls are useful here to help create more pleasantly discrete-sounding sonification and the controls of attack, decay, sustain, and release (ADSR) is partially automated to configure their values depending on the nature of sonification being discrete or continuous. The attack is set to 0.01 by default to prevent potential clipping at the start of a sound. If continuous, then the sustain is set to 1; If discrete, the sustain is set to 0 with the decay being 0.2, also shown in Figure \ref{fig:eeg_more}. These values could be further modified by the user interactively with sliders. If the user changes the default values, the values are saved when the user toggles between discrete and continuous. 

\subsection{Air Quality Data Sonification}

The air quality data is collected from the Canadian provincial websites for British Columbia, Alberta, and Ontario for the year of 2014. Sulphur dioxide (SO\textsubscript{2}), ozone (O\textsubscript{3}), nitrogen dioxide (NO\textsubscript{2}), carbon monoxide (CO), and particulate matter (PM\textsubscript{2.5}) were used as the metrics of air quality, and the data was sampled hourly.

\subsubsection{Replication of Sonification}

Like the original paper, each data attribute is configured to have 2 sine oscillators being the modulator and carrier for FM synthesis. The data is mapped to the modulation index (or the amplitude of the modulator), and also the amplitude (of the carrier) \cite{St_Pierre_Droumeva_2016} \cite{soundcloud_air}. The panning of the audio for each data attribute is configured in the audio visualization view by dragging the speaker to the left or right of the middle listener. The colours of the speakers are matched with those of the tracks. 

\begin{figure}[t]
\centering
\includegraphics[width=0.95\columnwidth]{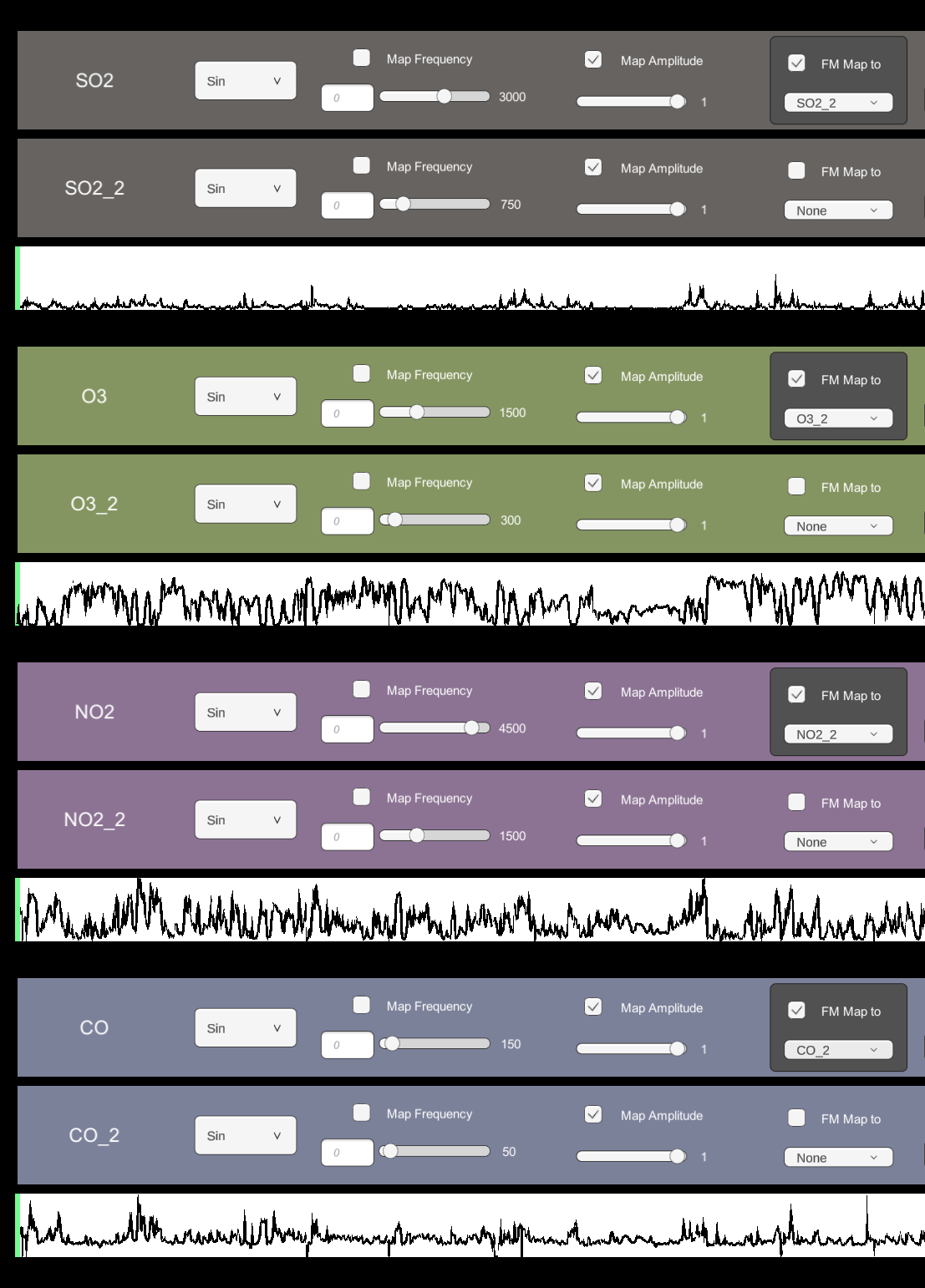}
\caption{The configuration to replicate the sound design of the sonification of air quality data, where the first row of each data attribute serves as the modulator, and the second is the carrier for FM synthesis.\label{fig:air_replicate}}
\end{figure}

\begin{figure}[t]
\centering
\includegraphics[width=0.95\columnwidth]{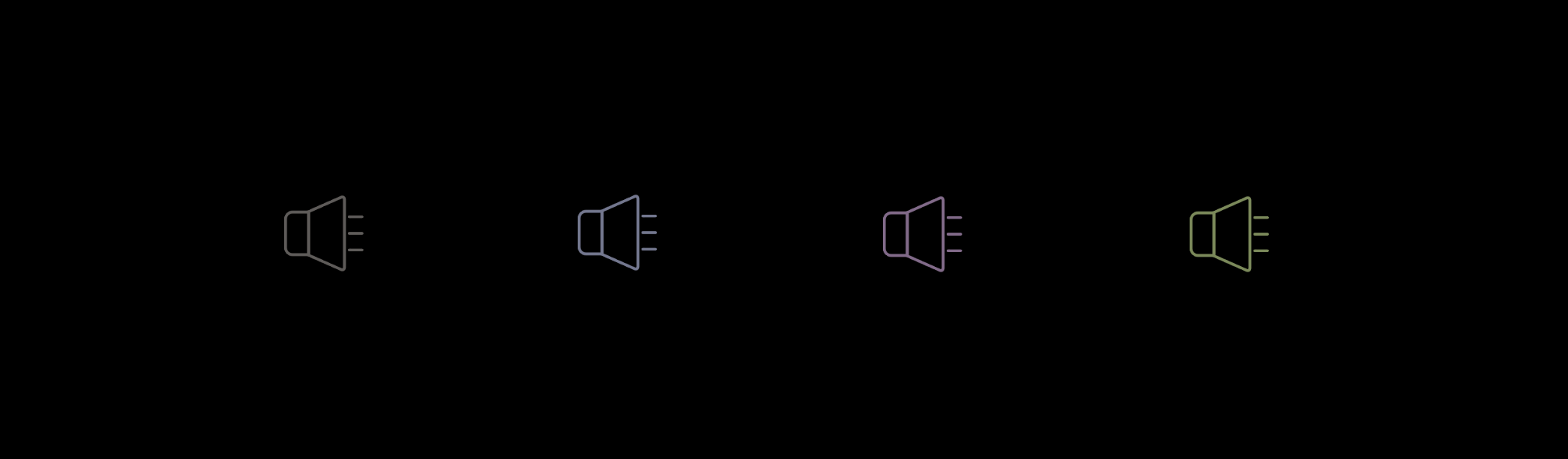}
\caption{In the audio visualization view, the panning of the sonification of air quality data could be replicated and configured. The center is the location of the listener. The differently coloured speakers refer to the data attributes with the corresponding colours, and their locations indicate the panning; For example, the green speaker refers to O\textsubscript{3} in Figure \ref{fig:air_replicate} and its panning location is on the far right. 
\label{fig:air_pan}}
\end{figure}

Since the exact configurations of FM synthesis used for SO\textsubscript{2}, O\textsubscript{3}, NO\textsubscript{2}, CO were not provided, we follow the general descriptions closely resulting in the following configurations. As shown in Figure \ref{fig:air_replicate}, for SO\textsubscript{2} the fundamental frequency is 750 Hz set for the carrier, and the carrier:modulator frequency ratio is 1:4, resulting in a frequency of 3000 Hz for the modulator; The panning is on the far left and configured in Figure \ref{fig:air_pan} by dragging the speaker of the corresponding colour of the data attribute and moving its location to the left or right for different panning. For O\textsubscript{3}, the fundamental frequency is 300 Hz and the modulator frequency is 1500 Hz with a ratio of 1:5; The panning is on the far right. NO\textsubscript{2} has a fundamental frequency of 1500 Hz and modulator frequency of 4500 Hz with a ratio of 1:3, and its panning is on the slight right. Lastly, CO has a fundamental frequency of 50 Hz and modulator frequency of 100 Hz with a ratio of 1:2, and its panning is on the slight left. The play rate is the same where each data point gets sonified for 0.2 seconds. Another attribute PM\textsubscript{2.5} is not sonified because granular synthesis was used but is not currently supported by our sonification framework. 


\pagebreak

\subsubsection{Achieving More}


The FM synthesis in the original paper was used to synthesize sounds specific to each data attribute. In our work, we could link the oscillators (being the modulator or the carrier) together without any restrictions and control them interactively in real-time, by mapping any oscillator to another target oscillator. This allows the data attaching the modulator to influence the sound texture of the carrier which could be attached to another data attribute. This is especially helpful to hear how one data stream affects another. For example, if we want to investigate the relationship between CO and NO\textsubscript{2} data, and if data attribute CO has a modulator (where the data is mapped to the amplitude) connected to a carrier of data attribute NO\textsubscript{2} (where the data is mapped to the frequency), the sonification sounds more metallic in its texture when the value of CO data gets larger (and vice versa), and sounds higher in its pitch when the value of the NO\textsubscript{2} data gets larger (and vice versa) to hear the potential effect of changes in CO levels on NO\textsubscript{2} levels. This is demonstrated in \href{https://youtu.be/yjWBIGvqRbU}{this video}\footnote{\url{https://youtu.be/yjWBIGvqRbU}}. More oscillators can be connected forming an algorithm to produce more complex sound involving multiple data attributes. (Sample selection is not supported because FM synthesis only applies to oscillators.) Moreover, the type of oscillators could also be chosen dynamically, instead of being restricted to only sine waves. 


Interleaved playback is also supported in our framework allowing easier comparison between data attributes. If this mode is turned on for the selected data attributes, the attributes are sonified one by one at each position in the dataset for the user to hear each data individually in an interleaving way. For example in this case if the data is mapped to the frequencies of the sound source, and the 4 pollutants are to be sonified using interleaved playback, we could hear 4 sounds played sequentially coming from the 4 attributes; The highest-pitched sound signifies the corresponding attribute has the highest value at this time. This then repeats for the next position in the dataset, as demonstrated in \href{https://youtu.be/QceUDfS8d4s}{this video}\footnote{\url{https://youtu.be/QceUDfS8d4s}}. Interleaved playback allows more exact and clearer comparisons, instead of hearing everything at once. 

\begin{figure}[t]
  \centering
  \includegraphics[width=0.95\columnwidth]{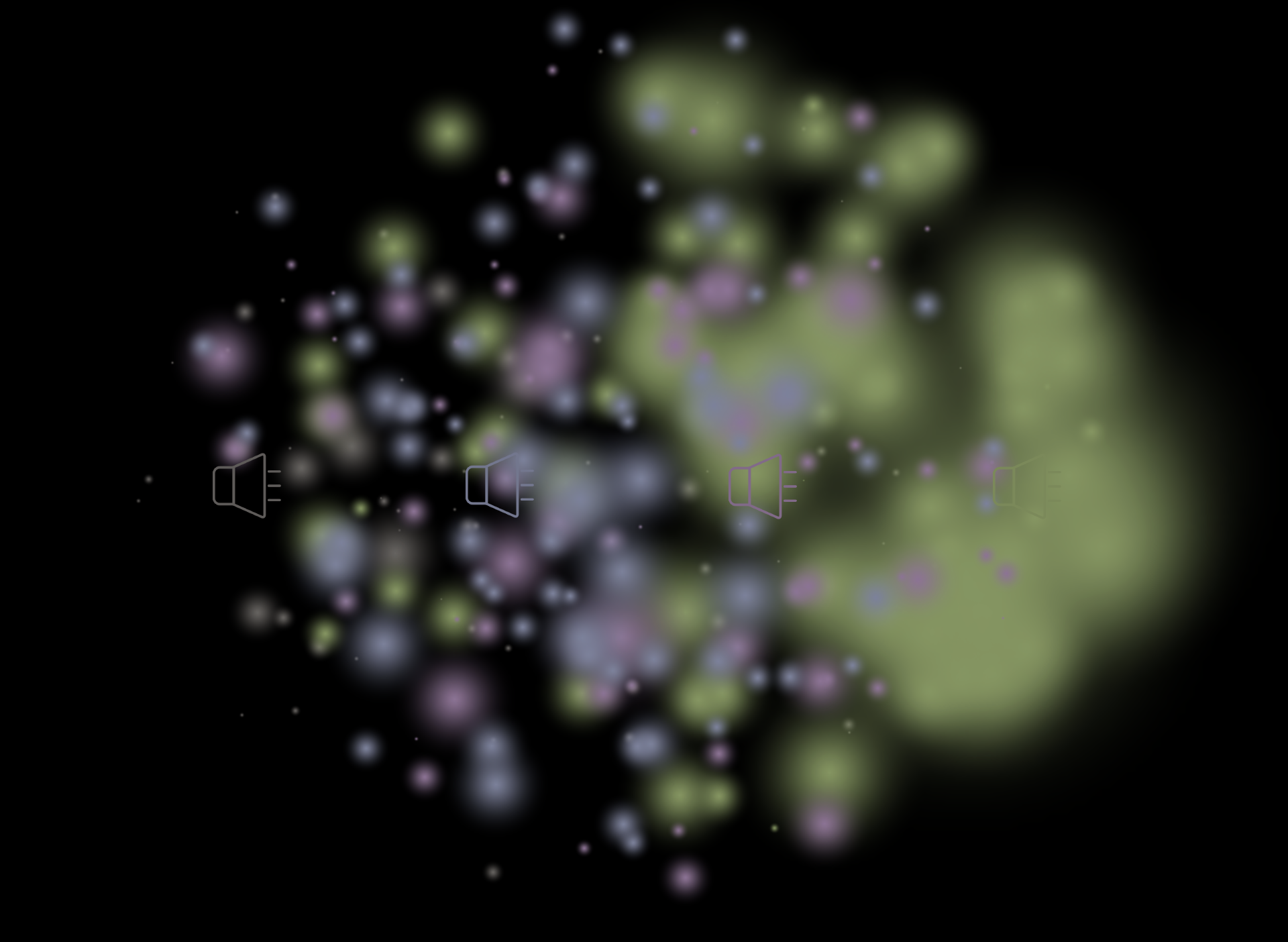}
\caption{This is the audio visualization view during the playing of the sonification of air quality data. At this time instance, green is the predominant colour meaning that O\textsubscript{3} is the most air pollutant.  
\label{fig:audio_vis}}
\end{figure}

In the audio visualization view of our framework, the user can drag and move the locations of the speakers to achieve and experiment with different audio pannings in real-time. Moreover, as seen in Figure \ref{fig:audio_vis} and \href{https://youtu.be/VVkg1i4y_ss}{this video}\footnote{\url{https://youtu.be/VVkg1i4y_ss}}, the audio is visualized with the particle system in Unity where the particles populated get more clustered when the audio gets louder or higher-pitched depending on the parameter mapping. As opposed to the visual plots shown in the track view, this aims to provide a more artistic and momentary visualization of the current audio instead of the data with the rendering power of Unity.

\section{Conclusion and Future Work} 
\label{sec:conclusion}


To conclude, this research project succeeds in developing a prototype that achieves various methods of sonification and visualization, improves the sonification experience of two datasets related to health and energy, and supports more functionalities. Future work includes support for more parameter controls and more synthesis methods allowing for more expressiveness of the sound; support for real time data input and more communication protocols such as Open Sound Control (OSC) \cite{Wright_Freed}, Serial and Musical Instrument Digital Interface (MIDI); being able to select the time scaling between data time and play time; being able to select regions of data to be played; 3D audio spatialization; and more comprehensive visualization. Moreover, it would be interesting to design interactive mechanisms within the framework to conduct certain tasks to evaluate the sonifications users create. Lastly, the framework itself should also be evaluated. This could be achieved by using case studies of various datasets sonified using the framework, or involving a user study asking users to perform tasks to evaluate the effectiveness of the framework. 


\begin{acknowledgments}
This research is supported by the Visual and Automated Disease Analytics (VADA) graduate program. 
\end{acknowledgments}

\bibliography{sonihed2022}

\end{document}